\documentclass[journal=jacsat,manuscript=article]{achemso}

\usepackage{achemso}
\usepackage{graphicx}
\usepackage{siunitx}
\usepackage{xcolor}
\usepackage{booktabs}
\usepackage{array}

\title{Dynamic Sensing via Photomodulated Gas Desorption in Plasmonic Nanoparticle Chemiresistors}

\author{Lukas Mielke}
\affiliation{Leibniz Institute of Polymer Research Dresden, Hohe Straße 6, 01069 Dresden, Germany}
\alsoaffiliation{These authors contributed equally to this work.}

\author{Dahee Heo}
\affiliation{Leibniz Institute of Polymer Research Dresden, Hohe Straße 6, 01069 Dresden, Germany}
\alsoaffiliation{These authors contributed equally to this work.}

\author{Gabriele Carelli}
\affiliation{Leibniz Institute of Polymer Research Dresden, Hohe Straße 6, 01069 Dresden, Germany}

\author{Hendrik Schlicke}
\affiliation{Leibniz Institute of Polymer Research Dresden, Hohe Straße 6, 01069 Dresden, Germany}

\email{schlicke@ipfdd.de}

\begin{document}
\newpage

\begin{abstract}
Quasi-static sensing, i.e., measurement of a sensor signal change induced by analyte presence  with respect to a baseline signal observed in a reference gas beforehand, is the state of the art for qualitative and quantitative analyte identification in chemiresistive sensing. 
However, this approach is prone to baseline drift and, especially for inexpensive miniaturized point-of-care sensors, repeated calibration with reference gas is not feasible. 
To access reliable baseline information, active switching between analyte and reference gas would be required, causing a limitation for miniaturized implementation.
Dynamic excitation of sensors offers an alternative: Via external stimuli the sensor is reversibly driven out of equilibrium in a controlled way, and the resulting transient sensor response, which is dependent on analytes present, is recorded and interpreted. Correlation of excitation and response signals can reduce effects of baseline drift and furthermore, dynamic features in sensor responses may contain valuable information.  
Especially light activation is a powerful and potentially miniaturizable approach to agitate sorption processes and therefore induce dynamics.
In this work we report photothermal heating of hybrid, chemiresistive gold nanoparticle composites via LED excitation matching the plasmon resonance, to dynamically shift the analyte sorption equilibrium and induce dynamic sensor responses in a highly controlled way. We demonstrate that these sensor responses enable the detection of volatile organic compounds (VOC) via interpretation of differential signal components and that this approach improves the chemiresistors' baseline stability.\\

\end{abstract}


\clearpage
\section{Introduction}
Hybrid materials composed of molecularly stabilized or cross-linked metal nanoparticles possess unique and highly tunable electronic, optical and chemical\cite{wessels_optical_2004} properties. The combination of charge transport characteristics, which are strongly influenced by sorption of volatile organic compounds (VOCs),\cite{steinecker_model_2007, joseph_gold_2008} widely adjustable chemical affinity\cite{liu_tuning_2024} as well as their time- and cost-efficient ink-based fabrication routes\cite{ketelsen_fully_2020} renders them promising for the fabrication of chemiresistive sensors. Their configurability allows for the fabrication of sensitive and selective electronic noses, when integrated as sensing arrays.\cite{jiang_confronting_2025} Especially the detection of VOCs in breath analysis is promising for diagnostics of diseases in a noninvasive manner.\cite{nakhleh_monolayer-capped_2014, harun-or-rashid_nanomaterial_2025}

Chemiresistive sensors are typically operated quasi-statically, meaning the sensing element's resistance, which is referenced to a baseline value is constantly or intermittently monitored and its changes indicate the presence of a target gas. However, sensor signals may be affected by instabilities or baseline drift, arising, e.g., due to environmental (such as temperature) changes, sensor contamination or irreversible changes to the active material due to aging.\cite{bulemo_selectivity_2025}. As a result, linking the measured quasi-static signal directly to a defined analyte concentration can be problematic. Such baseline instabilities, especially in nanoparticle composite chemiresistors that are usually operated at room temperature, pose significant hurdles when it comes to commercial applications.

Some of these challenges could be resolved by measuring differential responses, referenced to repeated exposures to a reference gas. However, given the technical complexity and strong miniaturization requirements of chemiresistive sensors, co-integration with reference gas systems is often not feasible. 

Another promising way to access differential sensor responses lies in triggering repeated, controllable and reversible shifts of chemical equilibria related to the sensing mechanism. For sorption-based chemiresistors, modulation of temperature is a feasible method to make use of thermodynamics and drive de- and adsorption processes.  
Since hybrid materials made from gold nanoparticles (GNP) exhibit pronounced plasmonic properties, their ability to strongly absorb visible light\cite{kholmicheva_prospects_2019,garcia_surface_2011} and efficiently convert it into heat\cite{jiang_size-dependent_2013,hartland_optical_2011,wang_understanding_2014} can be leveraged to dynamically modulate the temperature of GNP chemiresistors via pulsed light emitting diode (LED) or laser light irradiation. Here, perspectively co-integration with µLEDs is a  miniaturizable, technically straightforward approach.\cite{Vafaei2025,Prades2024}

Recently the beneficial use of plasmonic nanoparticles and micro-LEDs has been discussed in the context of metal oxide (MOX)-based gas sensing.\cite{suh_light-activated_2021,setka_photoactivated_2021} Advancing from sensors thermally activated via Joule heating, next generation light-activated sensors promise room temperature operation, a high degree of miniaturization and low energy consumption. 
Material design strategies include defects and doping, heterojunction formation and localized surface plasmon resonances. Here light absorption, charge engagement and surface activity can be used advantageously.\cite{lee_materials_2025} For example, Xu et al. reported enhanced gas sensing activity of ZnO nanotetrapods decorated with GNPs, attributed to their plasmonic properties.\cite{xu_light-activated_2018} Qomaruddin et al. utilized ZnO nanorods, decorated with electrophoretically deposited GNPs for plasmonically activated room-temperature NO$_2$ sensing.\cite{Qomaruddin2022} Cho et al. demonstrated improved irradiance and energy-conversion efficiency by integrating a monolithic $\mu$LED with a metal oxide nanowire gas sensor.\cite{cho_monolithic_2020} These examples highlight how efficient light absorption, miniaturization and integration capabilities of LEDs can be leveraged to enhance chemiresistive gas sensing. Additionally, nanojunctions consisting of GNP assemblies with sensing volumes down to \SI{0.0025}{\square\micro\meter} can transduce analyte sorption and offer further possibilities for miniaturization.\cite{fu_dielectrophoretic_2016} However, these concepts offer considerably broader opportunities for the development of optically activated sensing platforms.\cite{schlicke_plasmonic_2024} Especially the usage of machine learning, convolutional neural networks, support vector machines, and artificial intelligence are powerful tools for pattern recognition and response analysis.\cite{harun-or-rashid_nanomaterial_2025}. The addition of kinetic feature analysis is promising for qualitative and quantitative analyte identification in interfering environments.\cite{ogbeide_inkjet-printed_2022}

In this work we demonstrate dynamic photothermal excitation of GNP composite chemiresistors and utilize the resulting de- and adsorption dependent differential sensor signals for analyte detection and reduction of baseline drift-induced detection errors. 
Particularly, 1-propanol, heptane, octane, toluene and water were probed at varying concentration with a chemiresistor transiently excited using LED light, and the contribution of analyte desorption to the sensors' photothermal responses was extracted. Additionally, the response of GNP chemiresistors to arbitrary, time-dependent analyte concentration profiles was monitored upon continuous pulsed excitation. By interpretation of dynamic features in the resulting sensing signal, the analyte concentration was tracked and showed an improved long-term stability compared to a reference signal obtained from quasi-static read out of a second sensor. 

\section{Results and Discussion}
\begin{figure}[H]
    \centering
    \includegraphics{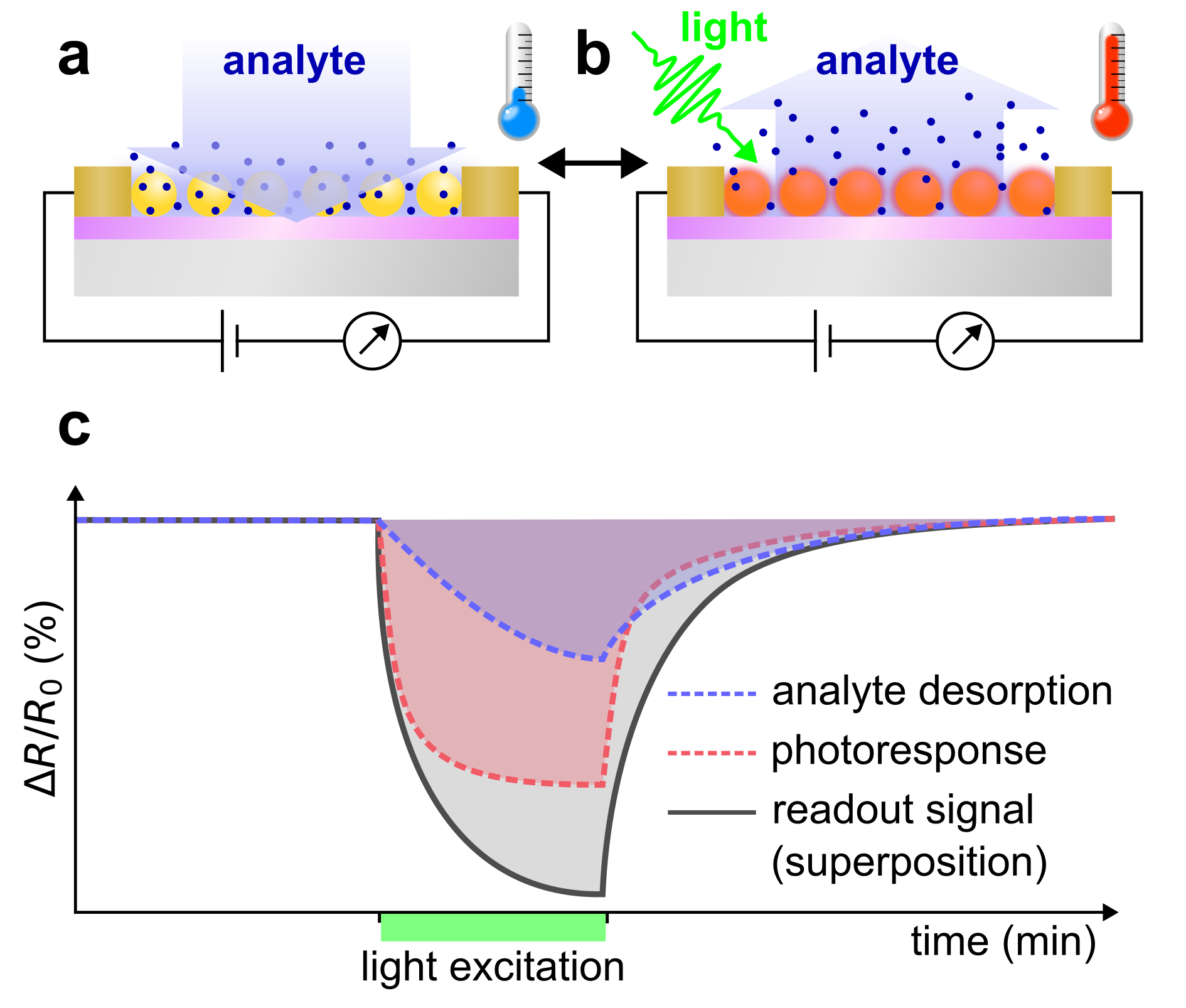}
    \caption{a) A GNP/ADT composite chemiresistor exposed to analyte gas and b) photothermal heating causing desorption. c) transient response signal to visible light excitation is a superposition of conductance changes due to photothermal heating and analyte desorption.}
    \label{fig:schematic}
\end{figure}

A schematic depicting the targeted dynamic sensing approach is shown in Figure~\ref{fig:schematic}. The resistance of a GNP-composite chemiresistor is constantly monitored while it is exposed to analyte target gas. Upon exposure, analyte molecules are sorbed by the sensing material, causing a significant, detectable change in sensor resistance (Figure \ref{fig:schematic}a).
The sensitive charge transport in GNP composites is typically well-described using a semi-empirical exponential model (equation \ref{eq:transport}).\cite{terrill_monolayers_1995}

\begin{equation}
    \sigma = \sigma_0 e^{-\beta\delta} e^{-\frac{E_\mathrm{A}}{kT}}
    \label{eq:transport}
\end{equation}

The first exponential term relates to tunneling-based inter-particle charge transport processes, with $\beta$ and $\delta$ denoting the tunneling decay constant and the interparticle spacing in the composites, respectively. The second term describes the thermal activation of transport, with $E_\mathrm{A}$ denoting an activation energy, that is commonly in the meV range and attributed to charging effects.\cite{zabet-khosousi_charge_2008} Sorption of analytes primarily causes swelling of the composite material, which increases $\delta$ and is typically reflected in a resistance increase.

While being exposed to the analyte, using an LED, the chemiresistor is transiently illuminated using visible light, which is converted into heat via thermalization of the excited localized surface plasmon resonance of the GNPs.\cite{mangold_surface_2009} The resulting temperature change shifts the analyte sorption equilibrium, inducing desorption of analytes from the sensing material (cf. Figure \ref{fig:schematic}b). Turning off the illumination again, the chemiresistor returns to its initial state by cooling to ambient temperature and re-sorbing the analyte gas.

Due to the thermally activated nature of the materials' charge transport, both analyte desorption and thermal activation contribute to the decrease of the materials resistance, resulting a readout signal that is a superposition of both effects (cf. Figure \ref{fig:schematic}c). 
To isolate the contribution of analyte desorption specifically, reference photoactivation measurements were conducted without the presence of any analyte gas, thereby calibrating the sensor.

\begin{figure}[H]
    \centering
    \includegraphics[width=\textwidth]{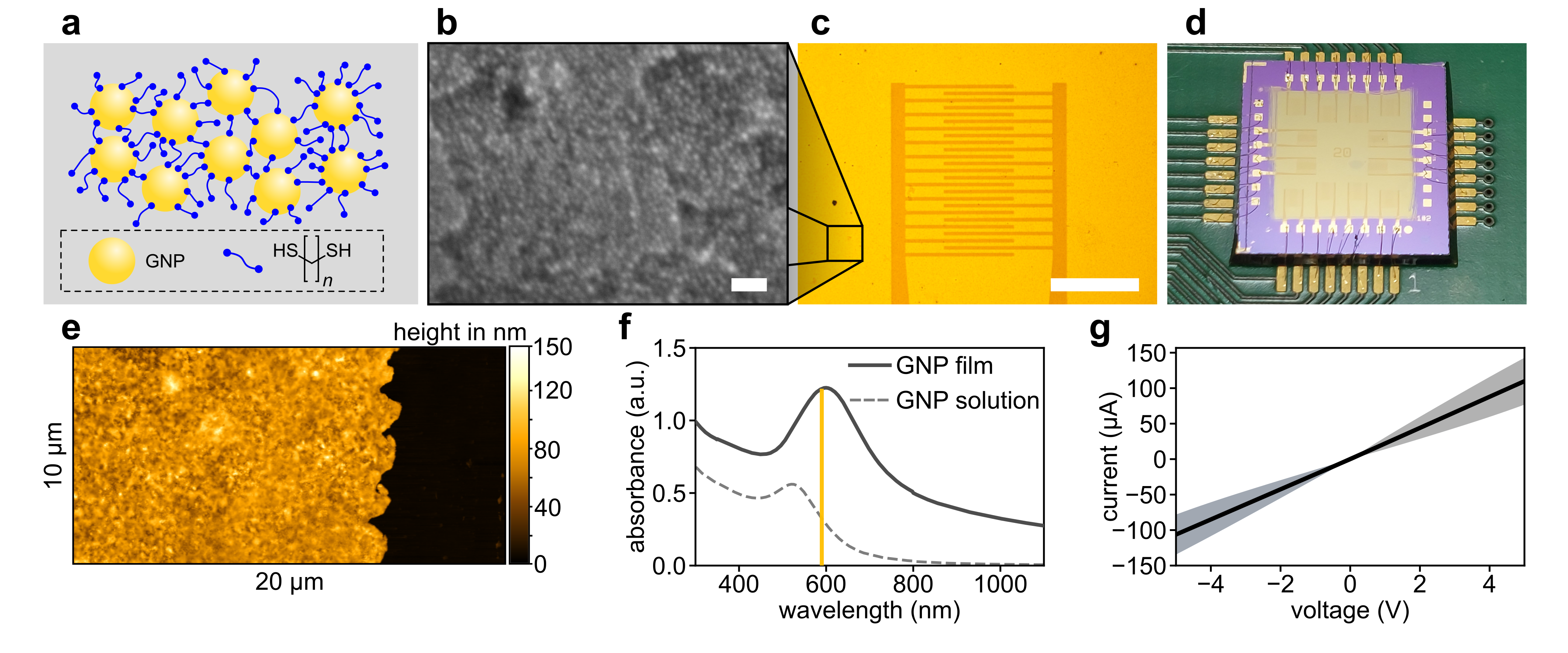}
    \caption{a) Schematic representation of a film containing gold nanoparticles and bifunctional molecular cross-linkers, i.e., alkanedithiols. b) Scanning electron micrograph of a 1,9-nonanedithiol cross-linked gold nanoparticle film (scale bar corresponds to \SI{40}{\nano\meter}). c) Microscope image of the film deposited onto interdigitated electrodes (scale bar corresponds to \SI{500}{\micro\meter}). d) Photograph of a chip equipped with 12 interdigitated electrodes, coated with an GNP/ADT composite film, mounted and wire bonded onto a printed circuit board. e) Topographic AFM image of the film fabricated by applying 5 deposition cycles, recorded at the edge of an deliberately introduced trench. A film thickness of 69~±~3~nm was determined. f) UV/vis absorbance spectra of the dodecylamine functionalized gold nanoparticles in solution (dilution factor f = 1/500 in heptane, optical path length d = 10 mm) and a GNP/9DT composite film. g) Current-voltage (IV) characteristics of 12~IDEs on one sensor chip. The solid line shows the current averaged over 12 devices, while the dark area depicts the standard deviation.}
    \label{fig:characterization}
\end{figure}

\paragraph{Device Fabrication and Sensor Material Characterization}
GNP chemiresistors were fabricated following an efficient layer-by-layer spin coating approach,\cite{schlicke_freestanding_2011} as described in detail in the Experimental Section. According to this procedure, a solution of 1-dodecylamine (12A) functionalized GNPs with an average diameter of $6.0\pm0.9~\mathrm{nm}$ and a solution of 9DT in methanol were alternatingly deposited onto an electrode microstructure, yielding the formation of thin film composed of a alkanedithiol (ADT) cross-linked GNP network. A representative TEM image of the nanoparticle solution, drop cast onto a TEM substrate and the corresponding size distribution can be found in the Supporting Information Figure~S1. Figure~\ref{fig:characterization}b shows a scanning electron microscope image of the GNP composite film and figure part c depicts a microscope image of the a film coated onto an interdigitated pair of electrodes (IDEs). In a typical device, the GNP film was contacted using 12 IDEs fabricated via standard photolithography processes on a Si/SiO\textsubscript{2} wafer (see the experimental section). Atomic force microscopy was conducted to determine the film thickness. A representative topographic AFM scan, conducted at a trench deliberately introduced into the thin film is shown in figure~\ref{fig:characterization}e. For this exemplary sample a thickness of 69~±~3~nm was extracted from the step profile. The observed granularity of the film is in accordance with previous reports.\cite{schlicke_freestanding_2015} 

Figure \ref{fig:characterization}f depicts absorbance spectra of a representative GNP film deposited onto a glass slide, fabricated alongside the device, as well as the GNP solution used for fabrication. The GNP film  shows a pronounced red-shift of the localized surface plasmon resonance (LSPR) maximum from 522~nm to 599~nm with respect  to the solution of 12A functionalized, isolated GNPs in heptane, which is attributed to plasmon coupling due to the proximity of the particles densely packed in the composite.\cite{brust_self-assembled_1998}
The film thickness was adjusted to ensure that incident light is majorly absorbed and converted into heat inside the GNP film. For the representative, typical film the absorbance at the plasmon peak was 1.23 which corresponds to absorption of $\sim 94~\%$ of the incident light in the film material. 

Following deposition, the sensor chip was mounted onto a carrier PCB and electrically contacted using wire bonding (Figure \ref{fig:characterization}e).
In order to calculate the conductivity of the GNP composite sensing material, current-voltage measurements were performed in the range of 5 to \SI{-5}{\volt} on the individual IDEs (cf. Figure \ref{fig:characterization}f). Slope fits were applied to extract the conductance. In combination with the derived film thickness and the geometry of the IDEs, a conductivity of \SI{6.5E-3}{\siemens.\centi\meter^{-1}}  was computed. This value is in agreement with values reported previously.\cite{schlicke_gold_2021}

\begin{figure}[H]
    \centering
    \includegraphics{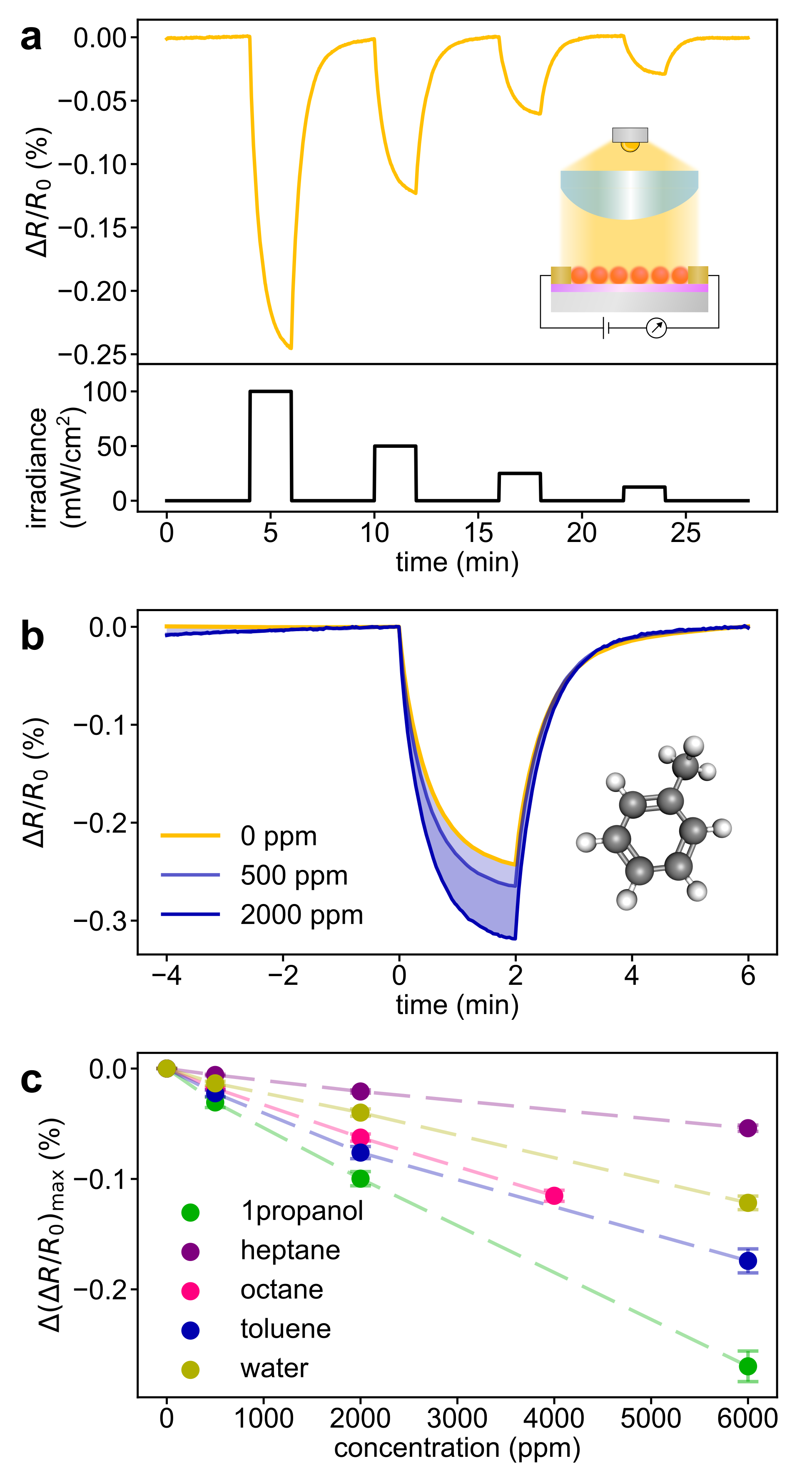}
    \caption{a) LED irradiance pulse sequence and resulting representative photoresponse of GNP chemiresistors. b) Pure photoresponse to 100~mW/cm\textsuperscript{2} of an amber colored LED (585-595~nm) and additional conductance change with 500 and 2000~ppm toluene. c) Maximum relative resistance change under analyte exposure, with difference relative to the maximum photoresponse at 100~mW/cm\textsuperscript{2} (585–595~nm, amber LED). 
    Dashed lines serve as guides to the eye.}
    \label{fig:results}
\end{figure}

The printed circuit board carrying the device was mounted inside of a custom-built measurement cell equipped with gas in- and outlets for introducing test gases and suitable for photoexcitation through a quartz window (Figure \ref{fig:results}a, schematic inset). 
An LED light source was mounted above the quartz window and adjusted to illuminate the sensor chip. A calibrated amber colored LED (wavelength range of 585 to 595~nm) was selected to match the spectral position of the LSPR maximum. During a typical measurement, the resistance of all 12 GNP chemiresistors on a chip was simultaneously and constantly monitored via a custom-built read-out circuit.

At first, the GNP film was repeatedly exposed with light pulses of the amber colored LED in pure nitrogen zero gas. Over the course of the excitation the pulse irradiance was varied between \SI{100}{\milli\watt.\centi\meter^{-2}} and \SI{12.5}{\milli\watt.\centi\meter^{-2}} (Figure \ref{fig:results}a). The exposure on-time was set to \SI{2}{min} while the off-time was set to \SI{4}{\min} to allow the device to recover fully. A representative photoresponse trace is shown in figure~\ref{fig:results}a. Observed relative resistance changes of the device ranged from -0.03 to -0.24~\% for an irradiance of 12.5 and \SI{100}{\milli\watt.\centi\meter^{-2}}, respectively. Note that even though these resistance changes are small, they can still be facilely resolved by conventional off-the shelf electronic components.

Subsequently, the device was exposed to different VOC vapors with concentrations of $500$ and $\SI{2000}{ppm}$. For each analyte application the resistive (dark) sensor response was allowed to saturate before the photoexcitation experiments were conducted. To probe photothermally induced desorption, each VOC concentration was applied three times and the exposure cycles shown in figure part a were also repeated thrice. Full time traces of the photoexcitation experiments are provided in the Supporting Information, section~S4 and data processing and evaluation are described in detail in the Supporting Information, section~S2. 
In brief, for every transient light exposure, a linear baseline fit was applied taking into account resistance data before and after exposure. The baseline data were then used to calculate the relative resistance change $\Delta R/R_0(t)$ upon light excitation. 

Resistance transients averaged over 11 devices and 3 exposure cycles are depicted in Figure \ref{fig:results}b. The amber colored curve represents the photothermally induced resistance change under zero gas (\SI{0}{ppm}), as shown in Figure \ref{fig:results}a. Here, for the given sensing element, a resistance change of -0.24~\% was observed. Repeated photoexcitations under exposure to \SI{500}{ppm} and \SI{2000}{ppm} toluene vapor yielded -0.26~\% and -0.32~\%. This well-resolvable difference in photoresponse, as indicated by the blue areas is attributed to photothermally induced desorption of toluene from the chemiresistors. The desorption contribution to the full signal is increasing with higher analyte concentration. 

To estimate the photothermally induced temperature change of the sensing film, the thermally activated charge transport model (equation \ref{eq:transport}) is considered. Rearranging the equation for a temperature dependent resistance change, equation \ref{eq:drr0-vs-T} is obtained:

\begin{equation}
\frac{\Delta R}{R_0} = e^{ \frac{E_A}{k_B} \left ( \frac{1}{T_{\mathrm{warm}}} - \frac{1}{T_{\mathrm{cold}}}\right ) }  - 1
\label{eq:drr0-vs-T}
\end{equation}

By rearranging equation~\ref{eq:drr0-vs-T}, the temperature change $\Delta T = T_\mathrm{warm}-T_\mathrm{cold}$ can be estimated from the resistance change due to illumination according to equation~\ref{eq:delta-T}. 

\begin{equation}
\Delta T
=
\frac{1}{
\frac{1}{T_{\mathrm{cold}}}
+
\frac{k_B}{E_A}\ln\!\left(\frac{\Delta R}{R_0}+1\right)
}
-
T_{\mathrm{cold}}
\label{eq:delta-T}
\end{equation}

For 9DT cross-linked GNPs between 3.2 and 4.0~nm, an activation energy of \SI{43}{meV} was reported.\cite{schlicke_tuning_2019} For larger particles of 7~nm, also cross-linked with 9DT, a value of $\sim$\SI{1200}{\joule/\mole} (or 12~meV) was observed,\cite{ketelsen_strain_2022} which is assumed to fit better to the particle system of this work with 6.0~nm mean diameter. 
With an ambient temperature of \SI{24}{\celsius} a temperature change of \SI{1.5}{\celsius} was calculated for \SI{100}{\milli\watt.\centi\meter^{-2}} exposure. Even though this value is low, it is sufficient to shift the sorption equilibrium of analytes inside the GNP/ADT composite. The temperature change could be significantly improved by a freestanding architecture of a film, where heat transfer to the substrate is reduced. For freestanding nanoparticle membranes. materials a temperature change of up to \SI{40}{\celsius} was observed but with laser light illumination of \SI{5.2}{\kilo\watt.\centi\meter^{-2}}.\cite{gauvin_plasmonic_2016}
 
Figure \ref{fig:results}c depicts the effective desorption contribution, i.e., the difference of photoresponses under zero gas and analyte gas $\Delta(\Delta R/R_0)_\mathrm{max}$ for exposures to 1-propanol, heptane, octane, toluene and water. The data was averaged from measurements on 11 individual chemiresistors on the sensing chip and 3 repetitions. 
The analyte dependence of the sensors' photoresponsivity matches the trends observed for chemiresistivity without photoactivation (see Supporting Information Figure~S3). Also the observed sensitivities widely agree with trends observed earlier for the given sensing materials\cite{schlicke_gold_2021}, however, a stronger response to 1-propanol was observed in this work. 

\begin{figure}[]
    \centering
    \includegraphics{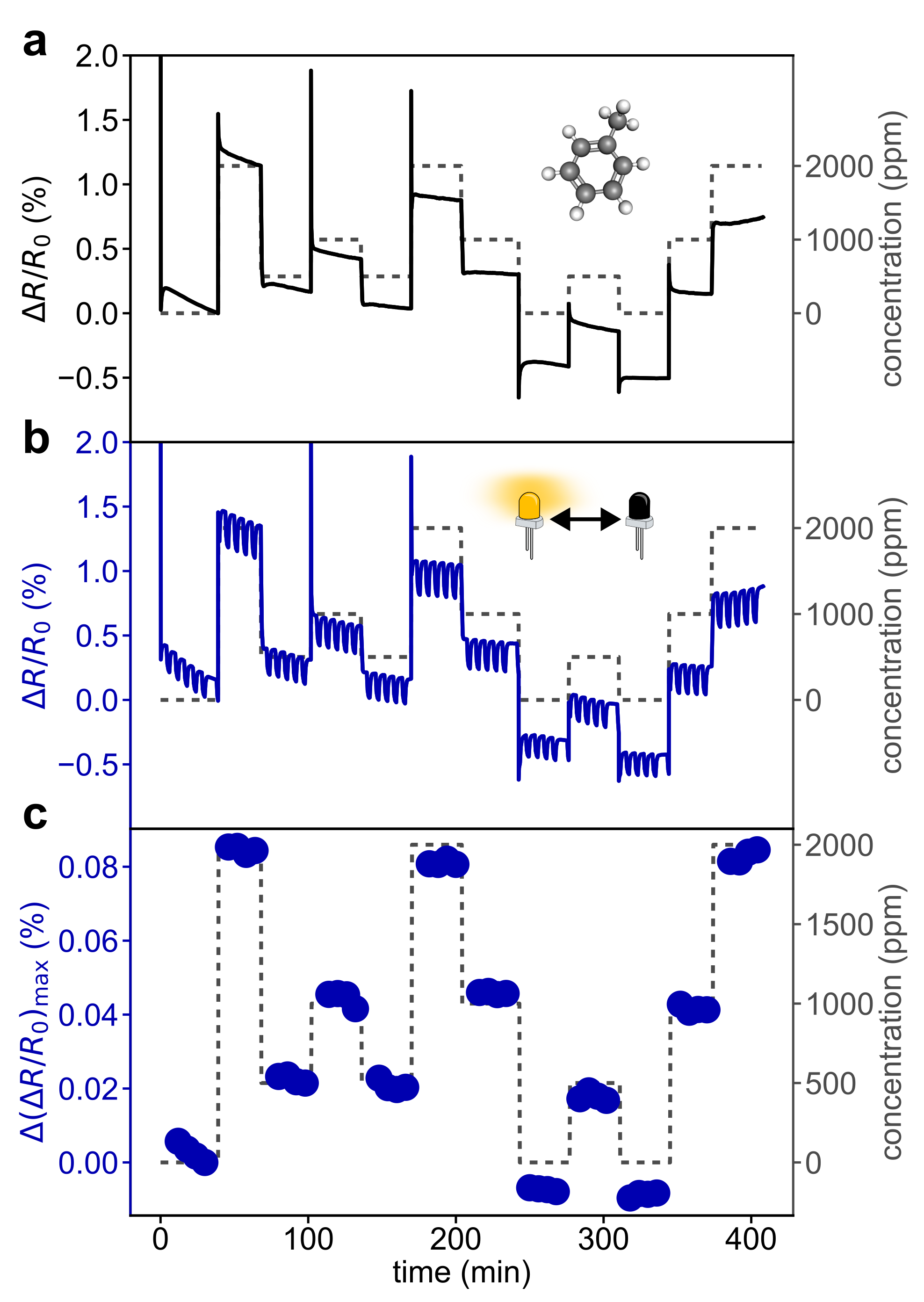}
    \caption{a) Representative chemiresponse normalized to the end of the first zero-gas (N\textsubscript{2}) exposure as reference value. The gray dotted line represents the actual set concentration profile. Note the increasing deviation because of baseline drift. b) Representative chemiresponse superimposed with continuous photoresponse to 100~mW/cm\textsuperscript{2} of an amber colored LED (585-595~nm). c) Maximum relative resistance change under analyte exposure, difference relative to the end of the first zero-gas (N\textsubscript{2}) exposure. The deviation towards the set concentration profile is noticeably lower.}
    \label{fig:baseline-comparison}
\end{figure}

\paragraph{Dynamic Readout}
To further emphasize the suitability of the discussed method for overcoming challenges of baseline drifts on in quasi-static chemiresistive measurements, we exposed a sensor to an arbitrary VOC concentration profile (toluene vapor) shown in figure~\ref{fig:baseline-comparison} on the second y-axis. Here, the vapor concentration varied between 0, 500, 1000 and 2000~ppm. The quasi-static chemiresistive response was recorded by monitoring the resistance of a reference chemiresistor placed in the test cell under dark conditions. The relative resistance change of the sensor is referenced to a first zero gas exposure (Figure part a). In parallel, we monitored the resistance of another chemiresistor that was repeatedly exposed to \SI{100}{\milli\watt.\centi\meter^{-2}} LED light with a wavelength of 585-595~nm and extracted the differential response features, i.e., the contributions of analyte desorption upon photothermal heating $\Delta (\Delta R/R_0)_\mathrm{max}$ as discussed above.

The relative resistance change observed in the quasi-static (dark) experiment depicted in figure part a, and the overall resistance change in the dynamic excitation experiment depicted in figure part b show a significant drift to lower resistance. Notably, while the first exposure to \SI{2000}{ppm} toluene vapor caused a resistance change of $\sim 1.14\%$, the second exposure to the same concentration was only $\sim 0.88\%$ and the third was $\sim 0.75\%$.
In contrast, taking into account the differential features of the sensor response introduced via pulsed excitation $\Delta(\Delta R/R_0)_\mathrm{max}$ as described above (Figure~\ref{fig:baseline-comparison}), a significantly reduced drift and an improved reproducibility of sensor responses is observed. This indicates that dynamic operation may be beneficial for improving the stability of GNP composite chemiresistors.
Differential sensing may be further used to reduce effects of baseline instabilities and enhancing sensitivity by exploiting lock-in techniques, i.e., phase-sensitive detection by correlating a faster-pulsed excitation with the measured response. 
While in this work, we only rely on interpretation of the amplitude of the dynamic sensor responses to photoexcitation, dynamic features in the response of GNP chemiresistors encode information on (e.g. sorption) process kinetics related to the chemical nature of the analytes, particularly their steric demand.\cite{schlicke_gold_2021}  Thereby, interpretation of the dynamic responses' shape can be further advantageous for analyte recognition.

\section{Conclusion}
We demonstrate that plasmonic photoexcitation of chemiresistors comprised of 9DT cross-linked GNPs can be used to reversibly trigger desorption of analytes and that resulting dynamic resistive responses can be employed for quantitative detection of VOCs. 

While pulsed photothermal excitation of GNP chemiresistors in pure nitrogen gas already induces a transient resistance decrease due to their thermally activated charge transport characteristics, the resistance decrease is significantly higher during exposure to analyte vapor, which was attributed to additional contributions of analyte desorption.
By interpreting these differential sensor signal changes introduced by photoexcitation, we successfully employed the dynamic sensing method for probing vapors of 1-propanol, heptane, octane, toluene and water vapor at concentrations between 500 and 6000~ppm. Furthermore, we demonstrated that the dynamic sensing scheme may be used to enhance the stability of the sensing signal.

\clearpage
\section{Experimental Section}

\textbf{Synthesis of Gold Nanoparticles} \\
Gold nanoparticles were synthesized following a procedure adapted from Peng et al.\cite{peng_facile_2008} \\
12A stock solution in n-hexane (4.3~mol/L) was prepared, of which 20~mL was preheated to 30~°C. Gold(III)chloride-trihydrate (99~mg, 0.25~mmol) was added and vigorously stirred for 30 min. TBAB (22~mg, 0.25~mmol) was dissolved in another 2.0~mL of the 12A stock solution. Subsequently it was rapidly added under vigorous stirring leading to the formation of gold nanoparticles, as indicated by an immediate color change. After 1~h 60~mL ethanol was added and thoroughly mixed. The resulting suspension was centrifuged for 5~min at 3745~g. The clear supernatant was discarded and the pellet was dried. Afterwards, the pellet was redispersed in heptane to the desired concentration and filtered through a 0.2~µm PTFE syringe filter. \\

\noindent
\textbf{Fabrication of 9DT Cross-Linked GNP Films} \\
1,9-Nonanedithiol (9DT) cross-linked films were fabricated following a layer-by-layer spin coating approach that was described by Schlicke et al.\cite{schlicke_freestanding_2011} In contrast to this procedure, we reduced the concentration of cross-linker to 2~mM giving an overall more homogeneous appearance of the film on the cm to mm scale.
The substrate was rotated at 300~rpm and solutions were added with a 20~s delay between depositions. First, 100~µL of the cross-linker solution (9DT, 2~mM, methanol) was applied twice. Then, 10~µL of the GNP stock solution and 20~µL of the cross-linker solution were alternatingly added until the desired thickness was reached. 
\\

\noindent
\textbf{Thin-film Characterization} \\
UV/vis absorbance spectra of GNP films were recorded using an Agilent Cary 5000 UV-Vis-NIR Spectrophotometer. Film thickness and surface roughness were evaluated using atomic force microscopy (AFM) using the ScanAsyst mode of a Dimension FASTSCAN atomic force microscope from Bruker Corporation. Conductivity values were calculated from current-voltage data shown in figure~\ref{fig:characterization}f recorded by using a Keithley k2602 and a k2604b source meter. \\

\noindent
\textbf{Fabrication of Microelectrode Structures} \\
Si/SiO\textsubscript{2} (500~nm) wafers were cut to the desired size and cleaned using acetone and IPA under ultrasonication for 3~min. Ti (10~nm) was evaporated using radio-frequency sputtering and Au (40~nm) using thermal evaporation with both processes carried out in a Korvus system. The substrates were baked at 200~°C for at least 10~min to removed adsorbed surface water. Negative AZ nlof 2020 photoresist was applied using spin coating (3000~rpm, 600~rpm/s, 60~s). The photoresist was soft-baked at 110~°C for 1~min and then exposed with UV-light using a custom-designed photomask and a Karl Suss MJB-3 UV 300 mask aligner for 3.5~s (16~mW/cm\textsuperscript{2}). This was followed by a post-exposure bake at 110~°C for 1~min and subsequent development in AZ 726 MIF for 1~min. The Au was selectively removed by an aqueous I\textsubscript{2}/KI/MQ etchant (mixing ratio 4:1:400, w/w/w). The residual photoresist was removed by immersion under vigorous stirring in TechniStrip~NI555 for 10~min at 70~°C. Lastly, the Ti was selectively removed by immersion in 5 w\% of triammonium citrate tribasic in 30\% H\textsubscript{2}O\textsubscript{2} at 35~°C. Before spin coating the substrates were hydrophilized using O\textsubscript{2} plasma for 3~min. \\

\noindent
\textbf{Photoexcitation and Analyte Vapor Measurements} \\
Photoexitation measurements were carried out using custom designed LED-mounts. LED intensities were derived from reference measurements using a calibrated Si photodiode. The GNP chemiresistors were placed in a custom-built measurement chamber (18~mL) and alternatingly exposed to light (100, 50, 25 and 12.5~mW/cm\textsuperscript{2}). This exposure series was repeated three times while all 12 electrode structures of the device were simultaneously recorded with a custom-built readout board. The readout circuit employs a shunt-based series configuration, in which the voltage drop at the node between the sensor and shunt resistor is sensed, amplified by an operational amplifier, filtered using an RC low-pass filter, and subsequently converted to a digital signal by an analog-digital converter. The off-times were set to be 4~min and the on-times were 2~min. As a reference, an identically fabricated device was measured in parallel under dark conditions. For photodesorption, gaseous analytes were provided in different concentrations (500, 1000, 2000, 4000 and 6000~ppm) until the GNP chemiresistor was saturated and subsequently the described photoexcitation series was performed. Each concentration was probed three times.

\clearpage
\section{Associated Content}
Supporting Information

\section{Author Information}
Corresponding Author \\
Hendrik Schlicke\\
*E-mail: schlicke@ipfdd.de \\

\noindent
ORCID \\
Lukas Mielke: 0009-0007-6002-0715 \\
Dahee Heo: 0009-0001-8568-0214 \\
Gabriele Carelli: 0009-0001-2451-4765 \\
Hendrik Schlicke: 0000-0002-6977-4042 \\

\noindent
Notes \\
The authors declare no competing financial interest.

\section{Acknowledgements}
H.S., and L.M. acknowledge ﬁnancial support by the German Research Foundation (Deutsche Forschungsgemeinschaft, DFG) via the research training group 2767 (Project-ID: 451785257). G.C. acknowledges financial support from the DFG via the “REC2” Cluster of Excellence (EXC3035, Project-ID: 533607596). We acknowledge support from our in-house scientific workshop for assistance with planing and building our measurement cell and readout boards. We thank Michael Göbel for support with scanning electron microscopy imaging.

\clearpage

\renewcommand{\thefigure}{S\arabic{figure}}
\renewcommand{\thesection}{S\arabic{section}}
\setcounter{figure}{0}
\setcounter{section}{0}

\begin{center}

\LARGE{\ \\[4cm]   {\sffamily \textbf{Supporting Information}}  }
\end{center}

\clearpage
\section{Characterization of Gold Nanoparticles}
\begin{figure}[H]
    \centering
    \includegraphics[width=\textwidth]{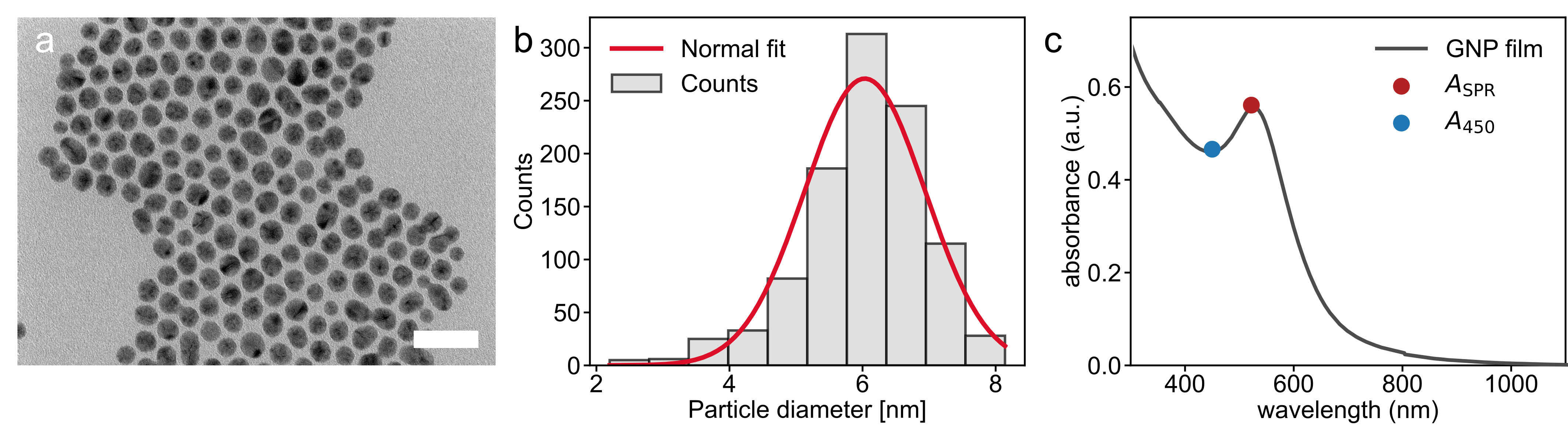}
    \caption{a) Representative transmission electron micrograph (scale bar: \SI{20}{\nano\meter}), b) corresponding size distribution histogram and UV/vis absorbance spectrum of the dodecylamine functionalized gold nanoparticles in solution (dilution factor f = 1/500 in heptane, optical path length d = 10 mm). The absorbance at $\lambda$ = \SI{450}{\nano\meter} A\textsubscript{450} used for concentration evaluation after Haiss et al. and the position of the maximum surface plasmon resonance A\textsubscript{SPR} is marked.\cite{haiss_determination_2007}}
    \label{fig-si:tem-histogram}
\end{figure}

The gold nanoparticle (GNP) batch used for film fabrication was synthesized according to the procedure of Peng et al. with slight modifications as described in the experimental section in the main document.\cite{peng_facile_2008} GNP sizes were determined by drop-casting the nanoparticle solution onto carbon-coated transmission electron microscopy (TEM) grids. A Talos F200i STEM was used. The particle diameters were calculated by extracting the projected area using the software Fiji (ImageJ 1.54p) and approximating a spherical particle. For analysis 1038 particles from 5 different grid positions were evaluated. Figure\ref{fig-si:tem-histogram} shows a representative transmission electron micrograph, the size distribution histogram of the GNP batch and the UV/vis absorbance spectrum of the solution in heptane, diluted by a factor of 500. The results are summarized in table~\ref{tab-si:tem-uv/vis}. The concentration was estimated via the UV/vis spectrum according to Equation~\ref{eq:particle_number} from Haiss et al.\cite{haiss_determination_2007} In the Equation $d$ corresponds to the mean particle diameter in nanometer extracted from the TEM images, $A_{450}$ is the absorbance at $\lambda = \SI{450}{\nano\meter}$, and $N$ is the number density of particles in \unit{\milli\liter^{-1}}.

\begin{equation}
N = \frac{A_{450} \cdot 10^{14}}{
d^2 \left[
-0.295 + 1.36 \exp\left(
-\left(\frac{d - 96.8}{78.2}\right)^2
\right)
\right]
}
\label{eq:particle_number}
\end{equation}

\begin{table}[ht]
\centering
\caption{Average particle diameter, LSPR maximum and estimated concentration.}
\label{tab:example}
\begin{tabular}{lccc}
\toprule
 diameter (nm) & $\lambda_{\max}$ (nm) & concentration (\,$\mu$M)\cite{haiss_determination_2007} \\
\midrule
 $6.0 \pm 0.9$ & 522 & 18\\
\bottomrule
\end{tabular}
\label{tab-si:tem-uv/vis}
\end{table}

\clearpage
\section{Data Processing and Evaluation}
\begin{figure}[H]
    \centering
    \includegraphics[width=\textwidth]{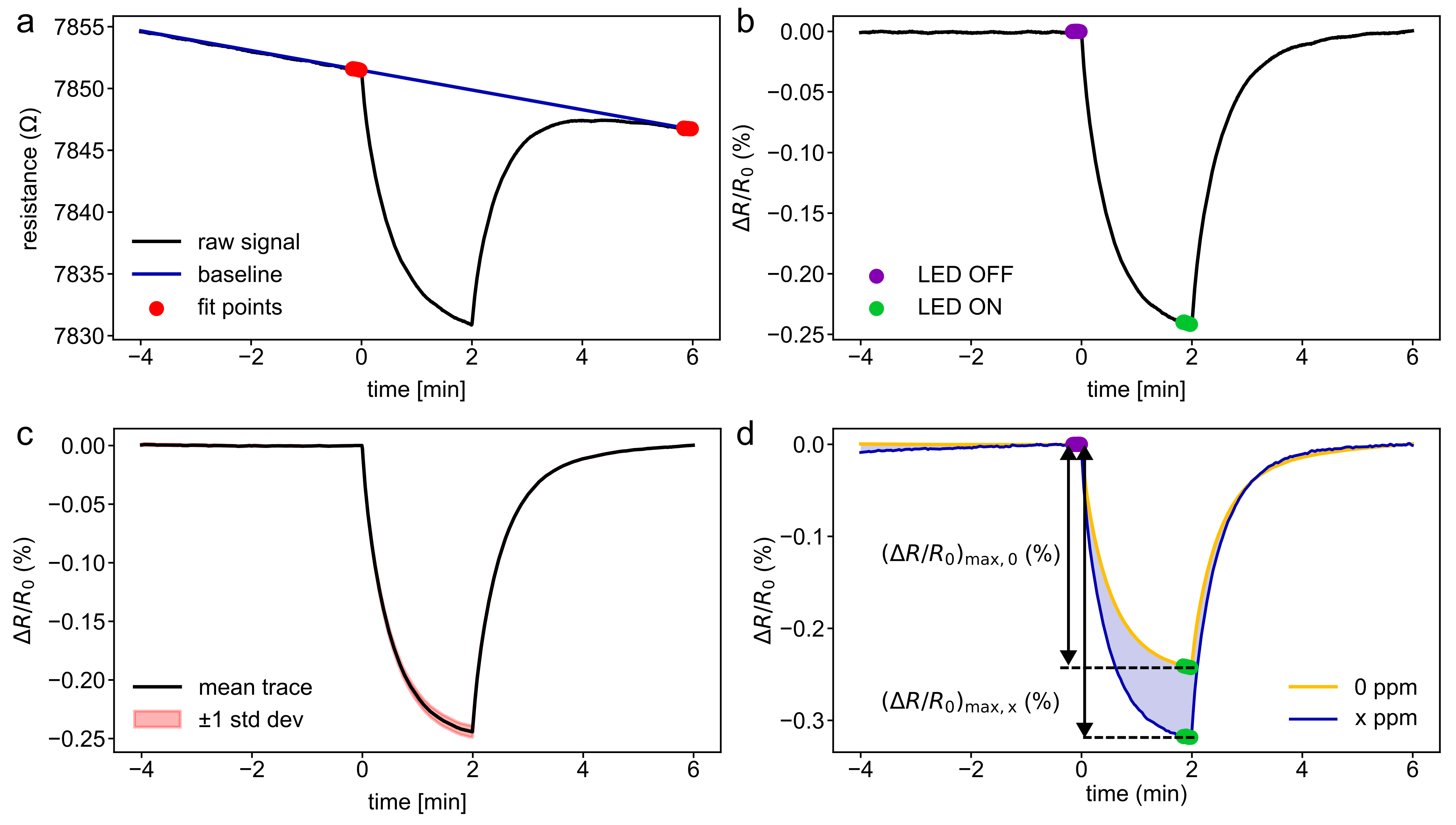}
    \caption{a) A range of \SI{10}{\s} before the transient photoresponse and after full recovery of the photoresponse was used for applying a linear fit and doing a baseline correction. b) A range of \SI{10}{\s} before the transient photoresponse and before stopping the transient exposure was averaged to define the maximum photoresponse. c) A mean trace of photoresponse of 11 interdigitated electrode pairs with standard deviation. d) An illustration of the definition of the $\Delta \left(\frac{\Delta R}{R_0}\right)_{\mathrm{max}}$ which related the maximum photoresponse under analyte exposure to the pure photoresponse.}
    \label{fig-si:photoresponse-evaluation}
\end{figure}

To evaluate the transient photoresponses in the presence and absence of analyte gas, the resistance dataset was segmented into sub-time-traces comprising \SI{4}{\min} pre-illumination dark resistance data, \SI{2}{\min} under illumination, and an additional \SI{4}{\min} in the dark. Note that “dark” refers to conditions without complete exclusion of ambient light. First, a baseline correction was performed by applying a linear fit to a \SI{10}{\s} interval of data points before the illumination and after full recovery of the signal (see Figure~\ref{fig-si:photoresponse-evaluation}a). The relative resistance change was calculated using Equation~\ref{eq:dR_over_R0} with $R_0(t)$ being the linear fit. 

\begin{equation}
\frac{\Delta R}{R_0}(t) = \frac{R(t)}{R_0(t)} - 1
\label{eq:dR_over_R0}
\end{equation}

\noindent
For every transient photoresponse the maximum photoresponse was calculated according to Equation~\ref{eq:dR_over_R0_max} by extracting points from a \SI{10}{\s} range before  illumination and before stopping the illumination (see Figure~\ref{fig-si:photoresponse-evaluation}b). 

\begin{equation}
\left(\frac{\Delta R}{R_0}\right)_{\mathrm{max}} =
\left(\frac{\Delta R}{R_0}\right)_{\mathrm{ON}} -
\left(\frac{\Delta R}{R_0}\right)_{\mathrm{OFF}}
\label{eq:dR_over_R0_max}
\end{equation}

\noindent
For Figure~3b of the main document the relative resistance changes from $11$ functional pairs of interdigitated electrodes on one chip were interpolated and a mean trace was calculated (see Figure~\ref{fig-si:photoresponse-evaluation}c. Within one measurement series each exposure to 100, 50, 25 and \SI{12.5}{\milli\watt.\centi\meter^{-2}} and 0, 500 and 2000~ppm analyte gas was repeated three times. The shown response traces in Figure~3b of the main document are a mean of those three repetitions. 

\noindent
Figure~3c of the main document shows the average maximum response for 11 pairs of interdigitated electrodes illuminated with an irradiance of \SI{100}{\milli\watt.\centi\meter^{-2}} upon exposure to 0, 500 and 2000~ppm of analyte gas. For every repetition the $\Delta \left(\frac{\Delta R}{R_0}\right)_{\mathrm{max}}$ was calculated according to Equation~\ref{eq:ddrr0max} by subtracting the maximum photoresponse from the previous 0~ppm exposure $\left(\frac{\Delta R}{R_0}\right)_{\mathrm{max},0}$ from the maximum photoresponse under analyte exposure $\left(\frac{\Delta R}{R_0}\right)_{\mathrm{max},x}$.

\begin{equation}
\Delta \left(\frac{\Delta R}{R_0}\right)_{\mathrm{max}} =
\left(\frac{\Delta R}{R_0}\right)_{\mathrm{max},x} -
\left(\frac{\Delta R}{R_0}\right)_{\mathrm{max},0}
\label{eq:ddrr0max}
\end{equation}

\clearpage
\section{Comparison of Photodesorption\\ and Chemiresistivity in Dark}
\begin{figure}[H]
    \centering
    \includegraphics[width=\textwidth]{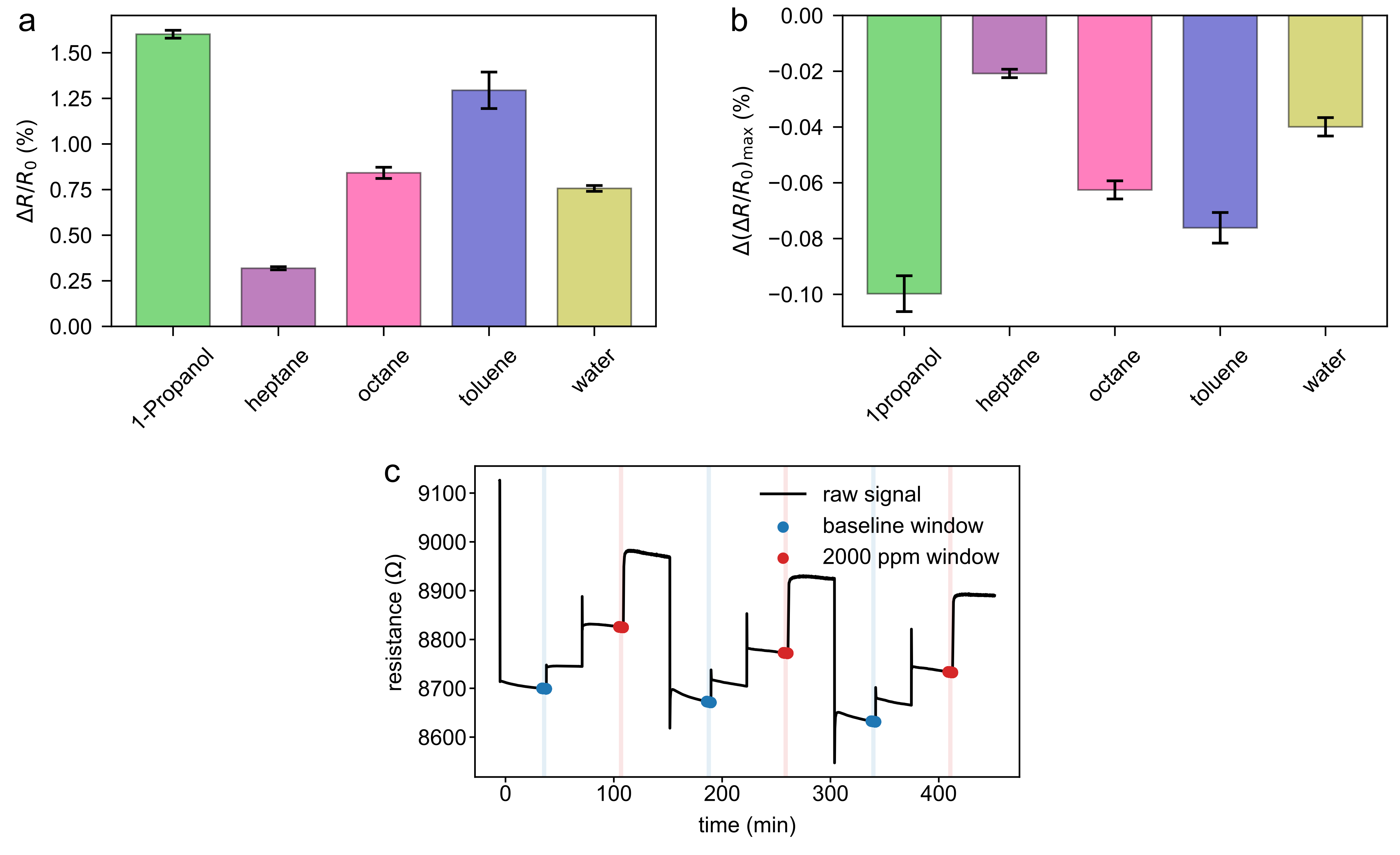}
    \caption{a) Chemiresistive response upon exposure to 2000~ppm of each analyte, extracted from 11 functional electrodes of a reference device measured in the dark simultaneously to the photoexposed device from the main document. Each exposure was repeated 3 times. b) Maximum relative resistance change due to photoexcited desorption  during analyte exposure (2000~ppm), 585–595~nm, amber LED. c) Exemplary raw signal from a reference device measured under dark conditions with \SI{4}{\min} windows used for $\frac{\Delta R}{R_0}$ calculations in a highlighted. }
    \label{fig-si:chemiresponse}
\end{figure}

The trends seen in Figure~3c in the main document for desorption of gaseous analytes were compared to the sorption of gas. For this, from a reference device measured in the dark, \SI{4}{\min} windows were defined in the end of the 0~ppm exposure and 2000~ppm exposure. The resistance values measured at \SI{0}{ppm} served as $R_0$ (see Figure~\ref{fig-si:chemiresponse}c). The relative resistance changes upon analyte exposure in the dark are shown in Figure~\ref{fig-si:chemiresponse}a. The analytes show similar sorption behavior like the desorption contributions shown in Figure~\ref{fig-si:chemiresponse}b.

\section{Response Traces of the Chemiresistors}
\begin{figure}[H]
    \centering
    \includegraphics[width=\textwidth]{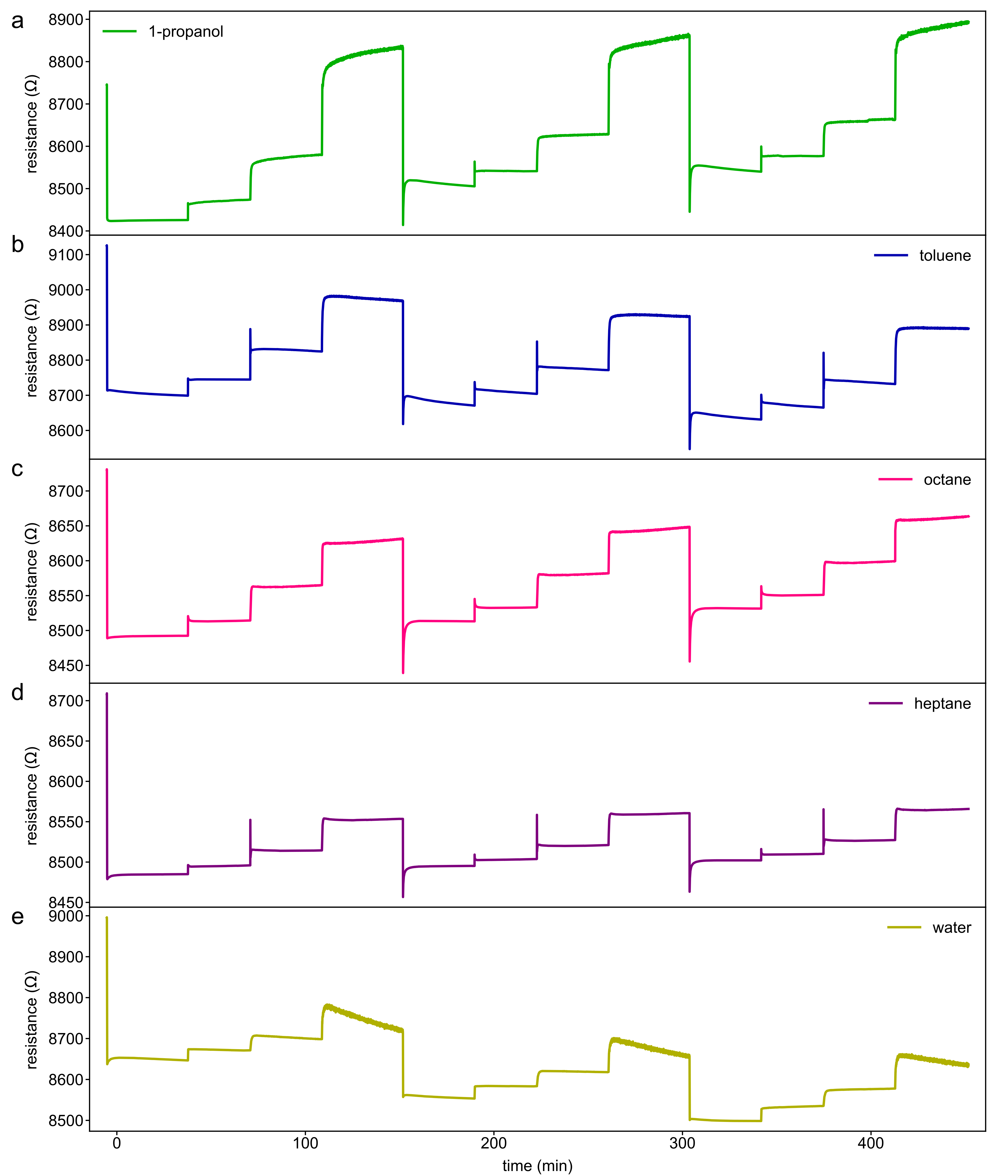}
    \caption{Exemplary response from a GNP/9DT chemiresistor to 0, 500, 2000 and 6000~ppm (note that octane exposure was only at a concentration of 4000~ppm instead of 6000~ppm) of a) 1-propanol, b) toluene, c) octane, d) heptane and e) water from a selected electrode.}
    \label{fig-si:raw-traces-dev010}
\end{figure}

\begin{figure}[H]
    \centering
    \includegraphics[width=\textwidth]{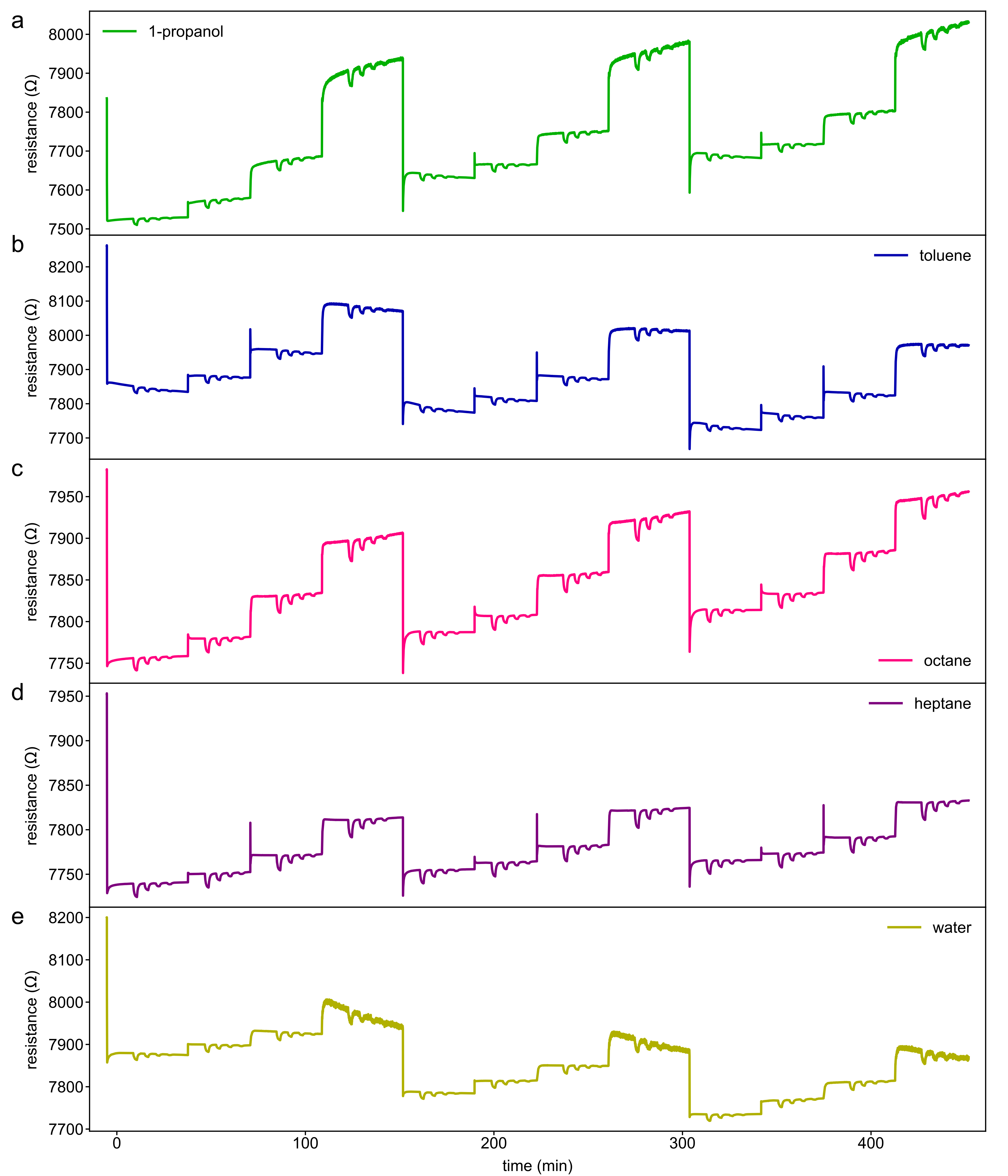}
    \caption{Exemplary response from a GNP/9DT chemiresistor to 0, 500, 2000 and 6000~ppm (note that octane exposure was only at a concentration of 4000~ppm instead of 6000~ppm) of a) 1-propanol, b) toluene, c) octane, d) heptane and e) water and additional transient photoexcitation to 100, 50, 25 and \SI{12.5}{\milli\watt.\centi\meter^{-2}} from a selected electrode.}
    \label{fig-si:raw-traces-dev009}
\end{figure}

Figure~\ref{fig-si:raw-traces-dev010} shows exemplary response data from a GNP/9DT chemiresistor to 0, 500, 2000 and 6000~ppm (4000~ppm for octane) of a) 1-propanol, b) toluene, c) octane, d) heptane and e) water. Figure~\ref{fig-si:raw-traces-dev009} shows exemplary response data from another GNP/9DT chemiresistor with additional photoresponse to 100, 50, 25 and \SI{12.5}{\milli\watt.\centi\meter^{-2}}.

\clearpage
\bibliography{references}

\end{document}